\documentclass[prb,preprint,aps,superscriptaddress,longbibliography]{revtex4-1}
\usepackage[]{color}
\usepackage{tabularx}
\usepackage{graphicx, epstopdf}
\usepackage{subfigure}
\usepackage{float}
\usepackage{amsmath}
\usepackage{amsfonts}
\usepackage{amssymb}
\usepackage{bm}
\usepackage{caption}
\usepackage{hyperref}
\usepackage{booktabs}
\usepackage{multirow}
\usepackage{svg}

\begin{document}

\title{Crystal Transformer Based Universal Atomic Embedding for Accurate and Transferable Prediction of Materials Properties}


\author{Luozhijie Jin}
\thanks{These two authors contributed equally to this work.}
\affiliation{School of Information Science and Technology, Fudan University, Shanghai 200433, China.}

\author{Zijian Du}
\thanks{These two authors contributed equally to this work.}
\affiliation{Department of Physics, Fudan University, Shanghai 200433, China.}

\author{Le Shu}
\affiliation{School of Information Science and Technology, Fudan University, Shanghai 200433, China.}

\author{Yan Cen}
\email{cenyan@fudan.edu.cn}
\affiliation{Department of Physics, Fudan University, Shanghai 200433, China.}

\author{Yongfeng Mei}
\affiliation{Department of Materials, Fudan University, Shanghai 200433, China.}

\author{Hao Zhang}
\email{zhangh@fudan.edu.cn}
\affiliation{School of Information Science and Technology, Fudan University, Shanghai 200433, China.}
\affiliation{Department of Optical Science and Engineering and Key Laboratory of Micro and Nano Photonic Structures (Ministry of Education), Fudan University, Shanghai 200433, China.}
\affiliation{State Key Laboratory of Photovoltaic Science and Technology, Fudan University, Shanghai 200433, China}


\date{\today}

\begin{abstract}
In the process of accelerating the prediction of properties of crystalline materials using machine learning methods, it is typically necessary to first embed crystal information to input the crystal structure. The method of generating embedding vectors for each atom significantly impacts prediction accuracy. Traditional practices involve either manually constructing atomic embeddings with the help of atomic property databases or obtaining atomic embeddings through training graph neural networks. However, these conventional methods do not significantly enhance the accuracy of property predictions. Our study focuses on the improvement of material property prediction through atomic embedding. We propose a novel approach to generate universal atomic embeddings, significantly enhancing the representational and accuracy aspects of atomic embeddings, which ultimately improves the accuracy of property prediction. Moreover, we demonstrate the excellent transferability of universal atomic embeddings across different databases and various property tasks. Our approach centers on developing the CrystalTransformer model. Unlike traditional methods, this model does not possess a fundamental graph network architecture but utilizes the Transformer architecture to extract latent atomic features. This allows the CrystalTransformer to mitigate the inherent topological information bias of graph neural networks while maximally preserving the atomic chemical information, making it more accurate in encoding complex atomic features and thereby offering a deeper understanding of the atoms in materials. In our research, we highlight the advantages of CrystalTransformer in generating universal atomic embeddings through comparisons with current mainstream graph neural network models. Furthermore, we validate the effectiveness of universal atomic embeddings in enhancing the accuracy of model predictions for properties and demonstrate their transferability across different databases and property tasks through various experiments. As another key aspect of our study, we discover the strong physical interpretability implied in universal atomic embeddings through clustering and correlation analysis, indicating the immense potential of our universal atomic embeddings as atomic fingerprints. For example, when we employed our universal atomic embeddings for the challenging task of predicting the formation energy in perovskite crystals, a task marked by data scarcity, we observed a significant enhancement in predictive accuracy. This outcome demonstrates that the robust transferability of our universal atomic embeddings not only ensures reliable accuracy in model predictions but also effectively counters the issue of underfitting in scenarios with limited data availability.
\end{abstract}

\flushbottom
\maketitle

\thispagestyle{empty}

\section{Introduction}
Materials science is currently undergoing a revolutionary transformation, driven by the integration of machine learning (ML) and artificial intelligence (AI). This integration is leading to unprecedented discoveries and deeper understandings of material behaviors and properties, previously inaccessible through traditional methods\cite{PredictingBandGaps}\cite{Butler2018}.
Central to this revolution is the nuanced interpretation and representation of atoms, the fundamental building blocks of materials. Traditional materials science emphasized the critical role of atoms in determining material properties and behaviors. The advent of ML and AI has added a new dimension to this understanding, offering innovative ways to represent and interpret atomic characteristics, thus defining a material's nature.
A key advancement in this field is the concept of atomic embeddings. These are sophisticated tools that translate the intrinsic characteristics of atoms into a mathematical form, offering a more detailed and holistic representation than simple numerical descriptors. These embeddings encapsulate not just the chemical identity of atoms but also their contextual role within the material's broader structure, marking a significant shift from conventional methods.

Our research introduces the CrystalTransformer model, a novel approach utilizing advanced ML techniques to generate universal atomic embeddings. Unlike traditional methods, CrystalTransformer learns a unique "fingerprint" for each atom, capturing the essence of their roles and interactions within the material. This model provides a new perspective for understanding material complexity and has the potential to set new standards in predictive modeling of materials.
The CrystalTransformer's universal atomic embeddings stand in contrast to existing graph neural network (GNN) models, which often provide limited representations of atomic interactions\cite{Xu2018PowerfulGNN}. Our model's embeddings are not only more comprehensive but also exhibit remarkable versatility and applicability across various contexts in materials science. This paper will explore the CrystalTransformer model's intricacies, its superiority over existing methods, and its efficacy in creating universal atomic embeddings that effectively serve as true fingerprints of atoms.

Furthering this innovation, our study extends the application of these universal atomic embeddings in multiple directions. We delve into the physical interpretability of these embeddings, employing clustering and correlation techniques to categorize atoms into distinct groups based on their electronic structures and chemical environments. This analysis enables us to establish a deeper connection between the embeddings and the intrinsic properties of atoms, solidifying their role as robust atomic fingerprints. Additionally, we examine the efficacy of these embeddings in reverse prediction of atomic properties, further demonstrating their potential in atomic-level material design and property prediction.Further, the excellent transferability of universal atomic embeddings signifies their capability to integrate knowledge learned from extensive databases in prediction tasks where data is scarce, thereby maintaining the model's generalization performance. For instance, we employed the CGCNN model as a baseline in the task of predicting the formation energy of inorganic perovskite materials. We observed a notable improvement in prediction accuracy upon applying universal atomic embeddings. This indicates that in certain critical tasks with limited data, the use of this technology can overcome the bottleneck of data insufficiency. It maintains the predictive capacity of the model even in scenarios of inadequate training, which is vital for accelerating the discovery of high-performance new materials.

In this work, our study not only highlights the importance of universal atomic embeddings in materials science but also establishes them as a cornerstone in constructing a unified language for atomic description. These embeddings, as innovatively developed in the CrystalTransformer model, transcend traditional boundaries, offering a deeper, more comprehensive understanding of material properties. This marks a significant advancement in the field, positioning universal atomic embeddings at the forefront of creating a standardized, insightful narrative for atomic characteristics and interactions.


\section{Results}

\subsection{The CrystalTransformer Model}

\begin{figure}[h]
    \centering
    \includegraphics[width=0.99\textwidth]{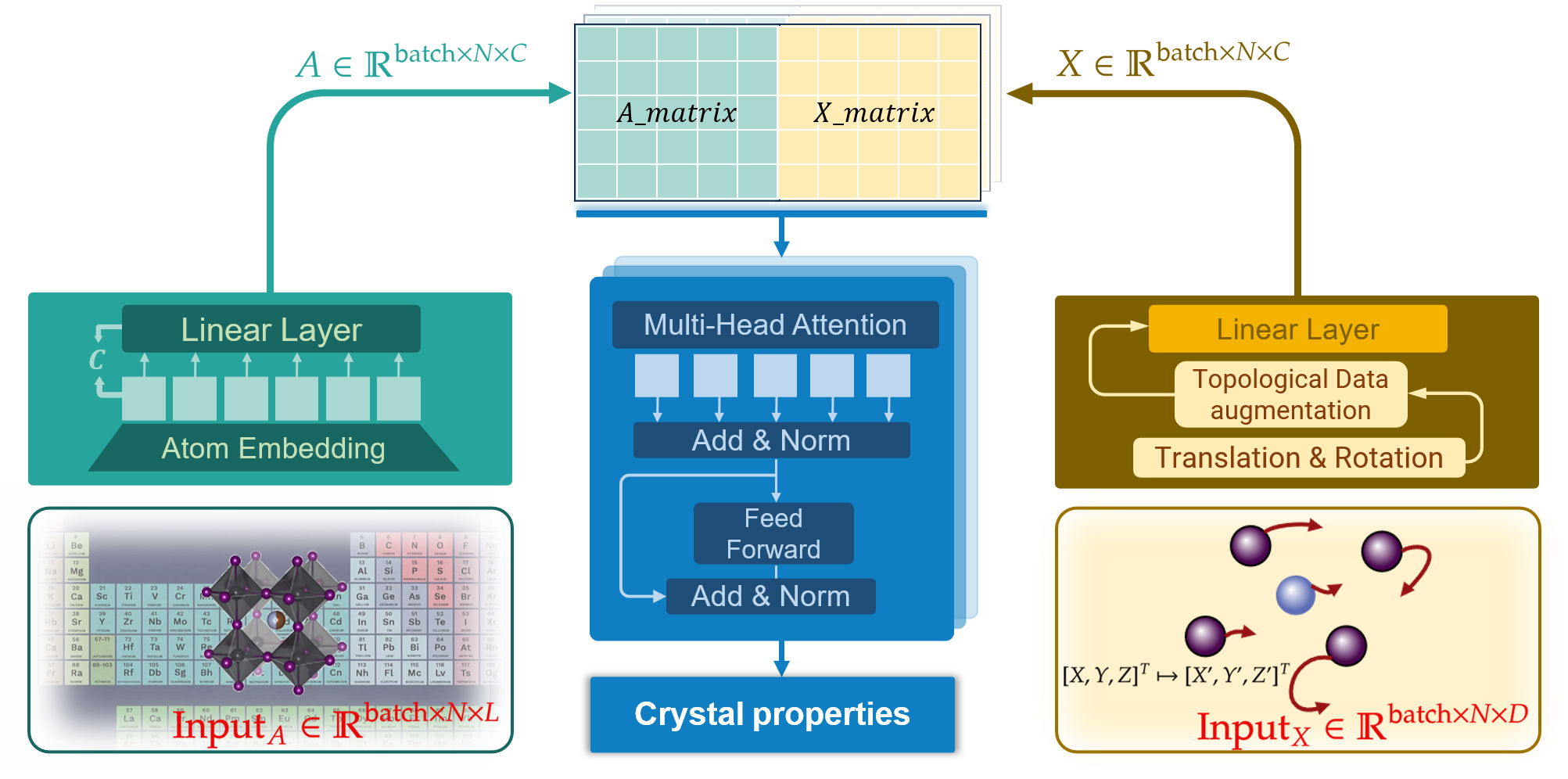}
    \caption{The structure of the CrystalTransformer model.}
    \label{fig:crystaltransformer}
\end{figure}

To build up the accurate model for the structure-properties relationship in materials, in this work, we introduce the vanillar transformer algorithm into the graph neural network, leading to the newly-developed crystal graph network algorithm based on the transformer called as the CrystalTransformer model, with the framework shown in Figure \ref{fig:crystaltransformer}. The CrystalTransformer model is designed to encode atomic features for crystal structure analysis. Given an input of atom features in a batch of size \( batch \), with \( N \) atoms per crystal, and \( L \) features per atom (where \( L \) includes a one-hot encoding of the atomic species), along with \( batch \times N \times D \) coordinates representing the spatial structure, with $D$ the spatial dimension. The model applies linear transformations to embed both feature and position information into a compatible dimensionality $C$, which can be described as follows,

\begin{equation}
A' = A W^{A} + b^{A},
\end{equation}

\begin{equation}
X' = X W^{X} + b^{X},
\end{equation}

where \( A \) is the \( batch \times N \times L \) tensor of atom features, \( X \) is the \( batch \times N \times D \) tensor of coordinates for atomic positions, \( W^{A} \) and \( W^{X} \) are the weight matrices for features and positions respectively, and \( b^{A} \) and \( b^{X} \) are their corresponding biases. The outputs \( A' \) and \( X' \) are both \( batch \times L \times C \) tensors.
These transformed features and position embeddings are then concatenated along the feature dimension as,

\begin{equation}
M = \text{Concat}(A', X'),
\end{equation}

where \( M \) is the concatenated feature matrix of shape \( batch \times N \times 2C \).
Then, the Transformer's encoder is applied to \( M \), which consists of multiple layers of self-attention and feed-forward neural networks, written as,

\begin{equation}
Z^{(l)} = \text{TransformerEncoderLayer}(Z^{(l-1)}),
\end{equation}

where \( Z^{(0)} = M \) and \( l \) indexes the layer of the encoder. Each Transformer encoder layer processes the input sequence and updates it through self-attention mechanisms and point-wise feed-forward networks, as described in the \ref{sec:transformer} section.
After processing the crystal structure features through the Transformer encoder, the CrystalTransformer model selects the first token from the output sequence for downstream prediction tasks. This token, representing an aggregated sequence representation, passes through a linear layer to produce the network's predicted material properties as,

\begin{equation}
\mathbf{y}_{\text{pred}} = \text{Linear}(Z^{(L)}_1),
\end{equation}

where \( Z^{(L)}_1 \) denotes the first token of the final encoder layer's output, and \( \mathbf{y}_{\text{pred}} \) is the material properties predicted by the network.
The CrystalTransformer's reliance on self-attention to process atom features and spatial coordinates allows it to learn representations that can capture the underlying physical interactions within the crystal. It should be noted that, this self-learned representation algorithm is advantageous over traditional graph neural networks like CGCNN, ALIGNN, MEGNET, and etc, which require predefined elemental features of constituent atoms and structural edge information such as bond lengths, bond angles, and Voronoi information as prior inputs. Hence, the CrystalTransformer described herein has the potential to encode atomic information more effectively, providing a robust approach for material property prediction, without the imposition of any structural priors.


To ascertain the robustness of the CrystalTransformer model, we curated a comprehensive set of well-developed open-source materials datasets. We mainly selected the Materials Project (MP)\cite{jain2013commentary} database, incorporated with the JARVIS-DFT\cite{choudhary2014jarvis} for a wider range of crystal material data and properties.
For MP\cite{jain2013commentary}, we meticulously chose the 2018.6.1 version to maintain comparability with other work, thus ensuring that our model's performance is benchmarked against established standards. This dataset contains 69,239 materials, including key properties like PBE bandgaps and formation energies, vital for assessing materials for different applications. For pretraining, we use the latest MP release (MP*), containing 134,243 materials.
The JARVIS-DFT\cite{choudhary2014jarvis} dataset, dated 2021.8.18, comprises 55,722 materials and a broad spectrum of properties crucial for the design of functional materials, which allows us to test the predictive power of our model across different material properties.

For training, validation, and testing splits, we adhered to established protocols for consistency. For the MP dataset, we followed the distribution of 60,000 (training), 5,000 (validation), and 4,239 (testing) as used in previous works. The MP* and JARVIS-DFT datasets and their properties were split into 80\% training, 10\% validation, and 10\% testing sets. 

\begin{table}[ht!]
\centering
\caption{Comparison of computational methods' MAE for predicting formation energy $E_f$\text{(eV/atom)} and bandgap $E_g$\text{(eV)}. The MP dataset contains 69,239 entries, whereas the MP\textsuperscript{*} dataset is the latest version with 134,243 entries.}
\label{tab:energy-bandgap-comparison}
\begin{tabular}{lcccc}
\hline
Method & MP-$E_f$ & MP\textsuperscript{*}-$E_f$ & MP-$E_g$ & MP\textsuperscript{*}-$E_g$\\
\hline
 CGCNN & 0.083 & 0.085  & 0.384 & 0.342\\
 MEGNET & 0.051 & \textbf{0.054} & 0.324 & 0.291 \\
 ALIGNN & \textbf{0.022} & 0.056 & \textbf{0.276} & \textbf{0.152}\\
 CrystalTransformer & 0.097 & 0.152 & 0.563 & 0.395 \\
\hline

\end{tabular}
\end{table}



In order to establish the CrystalTransformer model's accuracy on crystal datasets, we conducted comparative performance assessments against established graph neural network models. These models were evaluated on the MP dataset for formation energy ($E_f$) and bandgap ($E_g$), which are key properties for evaluating their thermal stabilities and optoelectronic performances. As listed in Table \ref{tab:energy-bandgap-comparison},
%
~despite lacking the prior inputs of atomic features and edge information of materials, the CrystalTransformer demonstrates competitive efficacy in predicting material properties, e.g. $1\sim4$ times larger in $E_f$ and $1\sim3$ times larger in $E_g$ compared to the traditional graph neural networks on MP/MP$^*$ datasets. This is partly because the Transformer architecture's self-attention mechanism provides enhanced generalization capabilities. Additionally, it does not strictly rely on predefined graph structures, but rather, it is capable of learning more extensive patterns directly from the crystal data. This underlines the model's robustness, particularly in capturing complex interactions within crystal structures through its self-attention mechanism.

\subsection{Universal Atomic Embeddings Method}

\subsubsection{Algorithm}

\begin{figure}[h]
    \centering
    \includegraphics[width=1\textwidth]{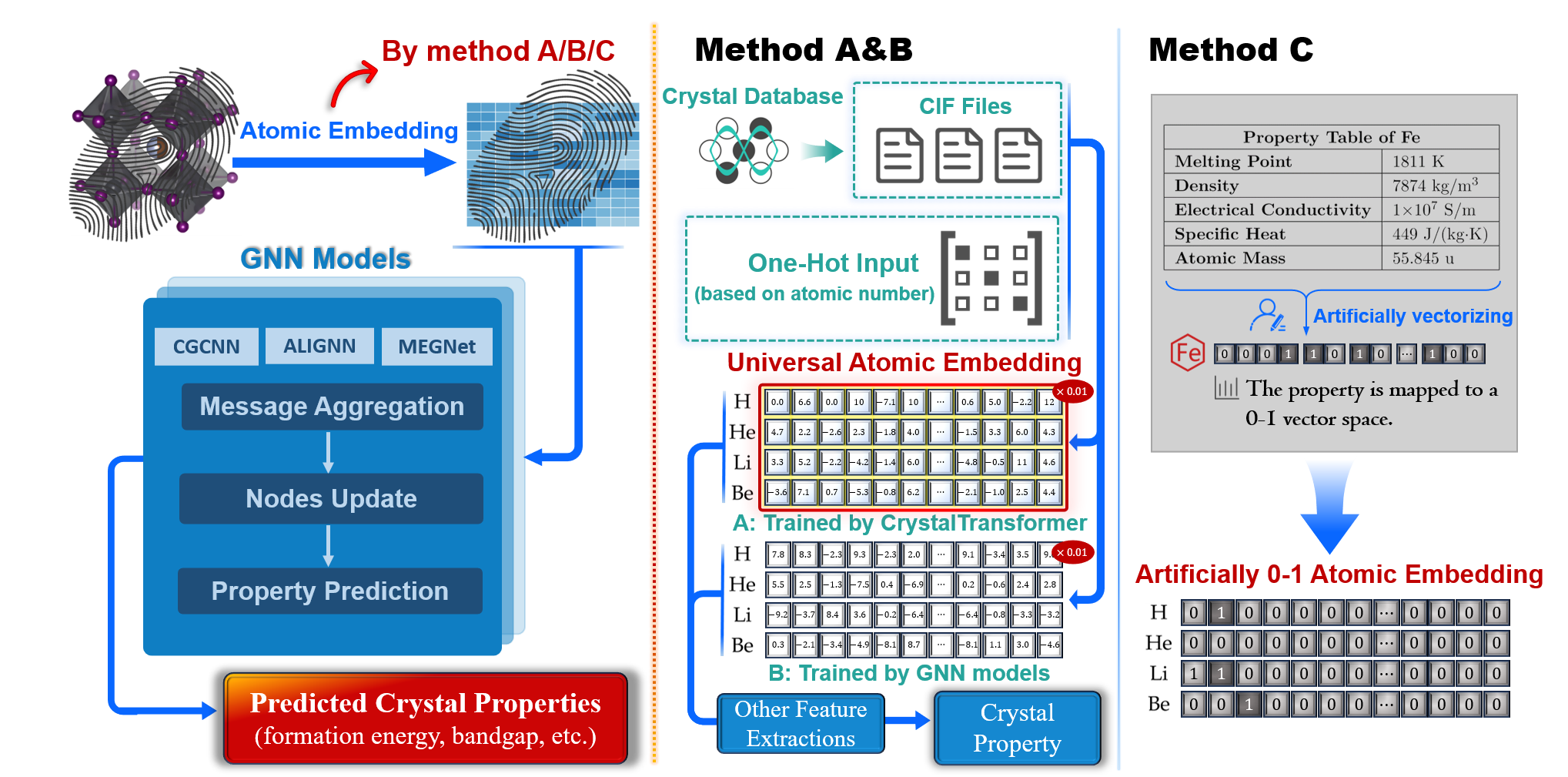}
    \caption{The generation methods and working principle of universal atomic embeddings.}
    \label{fig:theme}
\end{figure}

In subsequent experiments, we will demonstrate through a series of experiments that the atomic embeddings generated by the CrystalTransformer model are truly universal. In contrast, atomic embeddings produced by graph networks and manually constructed features lack this universality and do not significantly enhance the model's accuracy in property prediction. The process of predicting properties using atomic embeddings is illustrated in Figure \ref{fig:theme}. When predicting the properties of a material with known structural information and chemical composition, such as formation energy and bandgap, we first embed each atom in the crystal as node input information. This allows the encoding of crystalline materials into tensor forms that can be processed by the model. Then, we use mainstream GNNs such as CGCNN, ALIGNN, and MEGNet to perform deep feature extraction, including information transmission and aggregation, node feature updating, etc., and finally predict the crystal properties.

Method A involves training the CrystalTransformer model previously introduced in this paper and using the vector corresponding to each atom after the embedding operation as the atomic embedding. The CrystalTransformer model used in this method does not include the basic architecture of a graph network. This study will demonstrate that such embeddings are universal atomic embeddings with good transferability. Furthermore, this paper will use clustering methods and correlation analysis to show that universal atomic embeddings have potential physical interpretability, resembling atomic fingerprints in their capacity to reflect each atom's chemical information. In contrast, Method B involves training a graph neural network and using the vector corresponding to each atom after extracting the embedding as the atomic embedding. Both methods require the model to be extensively trained on crystal databases, such as the Material Project, and use the one-hot method to initialize the input vector to reduce prior knowledge.

Method C involves using a series of known atomic properties, such as atomic radius, density, and electrical conductivity, and arranging specific features into specific dimensions of a vector according to manually formulated rules. For example, in the atomic embeddings used in CGCNN, discrete atomic features are written into specific dimensions of the embedding using the one-hot method, and continuous atomic features are discretized into multiple intervals before being written into the embedding using the one-hot method, thereby obtaining atomic embeddings that explicitly reflect each atom's chemical information. The input features generated by this method are tensors containing only 0s and 1s. As a focus of this study, a series of subsequent experiments will prove that the atomic embeddings generated by the CrystalTransformer are superior to Methods B and C, and are universal atomic embeddings with high accuracy, good transferability, and potential physical interpretability, having the potential to become atomic fingerprints.

Following the initial validation of the CrystalTransformer model's performance in previous experiment, we meticulously examined the atomic embeddings from different models, i.e. CrystalTransformer, CGCNN, ALIGNN, and MegNet, which are first pre-trained on the expanded MP* dataset, focusing on the bandgap energy $E_g$ and formation energy $E_f$ predictive task. Subsequently, the embeddings are extracted and integrated into a CGCNN back-end model, trained on the original MP dataset. Table \ref{tab:222} show a comparative MAE analysis to determine the relative performance enhancements attributable to the front-end atomic embeddings.
Notably, when integrated into the back-end CGCNN model, these embeddings significantly reduce the MAE for both formation energy \(E_f\) and bandgap energy \(E_g\), compared to those without pre-training of atomic embedding as listed in Table~\ref{tab:energy-bandgap-comparison}. Among the atomic embeddings pre-trained by different models, those pre-trained by the CrystalTransformer model perform the best, with 14\% and 7\% reduction in $E_f$ and $E_g$, respectively. This improvement underscores the Transformer's superior capability in encoding complex atomic interactions, surpassing the conventional graph-based embeddings, which highlights the potential of Transformer embeddings in providing a more nuanced understanding of material properties, demonstrating their advantage in materials informatics.

\begin{table}[ht!]
\centering
\caption{Comparison of MAE results on the relative performance of different front-end atomic embeddings. Pre-training is conducted on the MP* dataset with $E_f$ (\text{eV/atom}) representing formation energy and $E_g$ (\text{eV}) indicating bandgap. Here, CT refers to CrystalTransformer, C to CGCNN, A to ALIGNN, and M to MEGNET, with the prefix denoting the front-end model and the suffix the back-end model.}
\label{tab:222}
\begin{tabularx}{\textwidth}{XXXXXXX}
\hline
Target & C-C & M-C &A-C & CT-C &CT-M &CT-A\\
\hline
$E_f$ & 0.074 & 0.082 &0.077 & 0.071 &0.049 & 0.018\\
$E_g$ & 0.378 & 0.457 &0.386 & 0.359 &0.304 &0.256\\
\hline
\end{tabularx}
\end{table}

Furthermore, our observations suggest that while the inductive biases of graph neural networks enhance their capabilities, they may also limit the scope of information captured in the atomic embeddings. Utilizing the CrystalTransformer model, which does not incorporate any predefined biases and relies solely on the model's self-learning capabilities for atomic representation, results in better and more transferable embeddings. This finding highlights the advantage of using a Transformer-based approach in generating embeddings that are not only effective in their primary tasks but also exhibit remarkable versatility and applicability in various material science contexts.


The integration of CrystalTransformer universal atomic embeddings into various graph neural network (GNN) models has led to quantifiable improvements in performance as listed in Table \ref{tab:222}.
Obviously, models like CGCNN, MEGNet, and ALIGNN were enhanced with these embeddings evaluated on the Materials Project (MP). The CGCNN model, when augmented with CrystalTransformer embeddings, denoted by \textbf{CT-C} in Table~\ref{tab:222}, shows a reduction in MAE values for formation energy \(E_f\), decreasing from 0.083 eV/atom to 0.071 eV/atom by 14\%, and for bandgap \(E_g\), decreasing from 0.384 eV to 0.359 eV by 7\%. Similar reduction can be observed for the MegNet, denoted by \textbf{CT-M} in Table~\ref{tab:222}, i.e. decreasing from 0.051 eV/atom to 0.049 eV/atom by 4\%, and for bandgap \(E_g\), decreasing from 0.324 eV to 0.304 eV by 6\%. And ALIGNN exhibits an improvement in \(E_f\) prediction accuracy as well, denoted by \textbf{CT-A} in Table~\ref{tab:222}, decreasing from 0.022 eV/atom to 0.018 eV/atom by 18\%, and for bandgap \(E_g\), decreasing from 0.276 eV to 0.256 eV by 7\%. 

\subsubsection{Cross-Task Transferability of Universal Atomic Embeddings} 

In addition to assessing the performance of embeddings pre-trained on one task, we further explore the cross-task transferability of universal atomic embeddings from the CrystalTransformer model by applying $E_f$-task-generated atomic embeddings to a different task, like bandgap prediction. The results, as listed in Table \ref{tab:mt_cgcnn}, reveal that these task-specific embeddings still enhance performance in other tasks, indicating their generalizability and underlying universal characteristics. For instance, the atomic embeddings trained for bandgap prediction, are used for formation energy tasks, leading to a measurable improvement in accuracy, with the MAE decreasing from 0.083 to 0.078 eV/atom by 6\% reduction. This finding not only highlights the versatility of the CrystalTransformer embeddings but also suggests a fundamental interconnectedness between different material property predictions, underlining the potential of these embeddings in diverse applications.



Inspired by the cross-task investigation, we further study the multi-task learning (MT) in the CrystalTransformer model, which exhibits a substantial impact on its performance when transferred to other models. As shown in Table \ref{tab:mt_cgcnn}, the embeddings derived from the MT-CrystalTransformer model, when transferred to the CGCNN model denoted as MT-CGCNN, yield superior performance compared to embeddings trained on individual tasks. The MT-CGCNN model achieves an MAE of 0.068 eV/atom for $E_f$ and 0.357 eV for $E_g$, outperforming the baseline CGCNN (by 18\% reduction in $E_f$ and 7\% reduction in $E_g$) as well as the CGCNN variants using single-task embeddings. This indicates not only the high quality and adaptability of the MT embeddings but also their enhanced transferability, proving their efficacy in diverse application scenarios.


\begin{table}[ht]
\centering
\caption{Comprehensive Performance Comparison: CGCNN Models on the MP Dataset with Different Embeddings. 'CGCNN' is the baseline model. 'CGCNN($E_f$)' represents the model with embeddings pre-trained on the formation energy task, while 'CGCNN($E_g$)' denotes embeddings pre-trained on the bandgap task. 'CGCNN(mt)' utilizes embeddings from the multi-task trained CrystalTransformer model. All models underwent pre-training on the MP* dataset and were further trained on the MP dataset.}
\label{tab:mt_cgcnn}
\begin{tabular}{ccccc}
\hline
Target & CGCNN & CGCNN($E_f$) & CGCNN($E_g$) & CGCNN(mt) \\
\hline
$E_f$ (eV/atom) & 0.083 & 0.071 & 0.078 & 0.068 \\
$E_g$ (eV) & 0.384 & 0.383 & 0.359&0.357 \\
\hline
\end{tabular}
\end{table}

However, the exploration of integrating coordinate embeddings alongside universal atomic embeddings within the CGCNN framework yields results beyond physical intuition. As listed in Table~\ref{tab:coords}, for formation energy $E_f$, the MAE increases from 0.071 eV/atom in the atom-embedding-only model to 0.085 eV/atom in the atomic-coordinate combined embeddings model. Similarly, for PBE bandgap $E_g$, the MAE worsens from 0.359 eV to 0.395 eV. This decline in performance suggests that while universal atomic embeddings are highly transferable and effective, the coordinate embeddings do not perform similarly. The result might indicate a limitation in the Transformer model's ability to effectively generalize structural information, highlighting an area for further development. However, it also underscores the potency and adaptability of atomic representations, emphasizing their central role in capturing transferable insights within the material-science domain.

\begin{table}[ht]
\centering
\caption{Performance comparison of CGCNN with with various embedding approaches in the CTS Framework. 'CGCNN($E_a$)' uses only universal atomic embeddings and is trainable , while 'CGCNN($E_a$+$E_c$)' combines atom and coordinate embeddings and 'CGCNN($E_a^f$)' uses frozen atomic embeddings which is untrainable. Pre-training was conducted on the MP* dataset, and models were trained on the MP dataset.}
\label{tab:coords}
\begin{tabular}{ccccc}
\hline
Target & CGCNN & CGCNN($E_a$) & CGCNN($E_a$+$E_c$)  & CGCNN($E_a^f$) \\
\hline
$E_f$ & 0.083 & 0.071 & 0.085 &0.073 \\
$E_g$ & 0.384 & 0.359 & 0.395 & 0.376\\
\hline
\end{tabular}
\end{table}

Furthermore, we also perform a compelling comparison of CGCNN models utilizing different embedding strategies. As listed in Table~\ref{tab:coords}, it is particularly noteworthy that, the performance of 'CGCNN($E_a^f$)', which employs frozen pre-trained embeddings from the CrystalTransformer, achieves an MAE of 0.073 eV/atom for formation energy $E_f$ and 0.376 eV for bandgap $E_g$, demonstrating only a slight increase in error compared to the fine-tunable 'CGCNN($E_a$)' variant, which records 0.071 eV/atom and 0.359 eV for $E_f$ and $E_g$, respectively. These results highlight the intrinsic value of the pre-trained embeddings themselves. 
The slight performance difference also underscores the potential benefits of fine-tuning, although the robustness of the frozen embeddings is evident, marking them as highly effective for material property predictions.

\subsubsection{Benchmark}

The efficacy of CrystalTransformer embeddings in enhancing the performance of the CGCNN model is further validated through modelling conducted on the JARVIS-DFT dataset. As indicated in Table \ref{tab:jarvis-performance}, the implementation of CrystalTransformer pre-trained embeddings (denoted as \textbf{CT-C}) demonstrates a notable improvement in predicting both formation energy $E_f$ and bandgap energy $E_g$.
Specifically, for the task of formation energy prediction, when employing CrystalTransformer atomic embeddings, the MAE is reduced from 0.135 eV/atom to 0.125 eV/atom by 7.7\%. Similarly, a more pronounced effect is observed in the prediction of bandgaps, where the MAE decreases substantially from 1.374 eV to 1.074 eV by 22.0\%, which underscores the proficiency of the pre-trained embeddings in capturing electronic bandstructures, a critical aspect for the exploration and development of new semiconductor materials.

\begin{table}[ht]
\centering
\caption{Performance comparison on the JARVIS-DFT dataset for formation energy ($E_f$) and bandgap ($E_g$). 'CGCNN' represents the baseline model, and 'CT-C' denotes the model using embeddings pre-trained with CrystalTransformer.}
\label{tab:jarvis-performance}
\begin{tabular}{cccc}
\hline
Target (JARVIS-DFT) & CGCNN & CT-C \\
\hline
$E_f$ (eV/atom) & 0.135 & 0.125  \\
$E_g$ (eV) & 1.374 & 1.074  \\
\hline
\end{tabular}
\end{table}



\subsubsection{Interpretability}

\begin{figure}[ht!]
    \centering
    \includegraphics[width=0.99\linewidth]{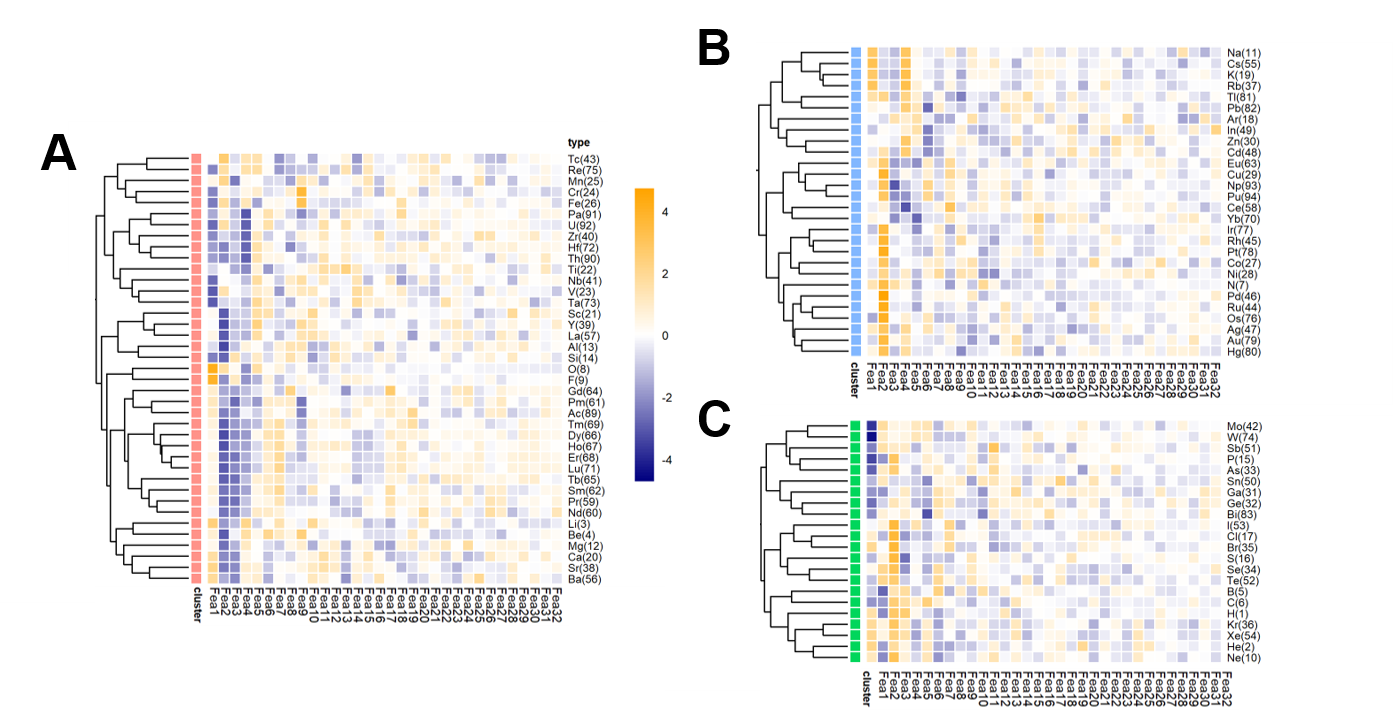}
    \caption{Cluster heatmap of atoms drawn based on 128 dimensional features}
    \label{fig:clusterheatmap}
\end{figure}

In order to elucidate the effectiveness of the atomic embeddings in reflecting the fundamental physical and chemical properties of atoms, we undertook an in-depth analysis using clustering and correlation techniques, to categorize these universal atomic embeddings into different groups, with each group revealing unique electronic structures, sizes, and potential chemical environments of the atoms. This clustering not only affirms the physical relevance of the atomic embeddings method developed in this work, but also provides a new perspective in understanding how elements influence the crystalline structure and material properties. The Principal Component Analysis (PCA) method is used to reduce the dimensionality of the data, thereby more effectively revealing the intrinsic relationships between atomic properties. Subsequently, the dimensions of the universal atomic embeddings are reduced from the original $89\times128$ to $89\times32$ dimensions, and the cluster heatmap, drawn based on these 32 variables, unveils potential structural patterns and interrelationships within the universal atomic embeddings, as shown in Figure~\ref{fig:clusterheatmap}. 

\begin{figure}[ht!]
    \centering
    \includegraphics[width=1.0\linewidth]{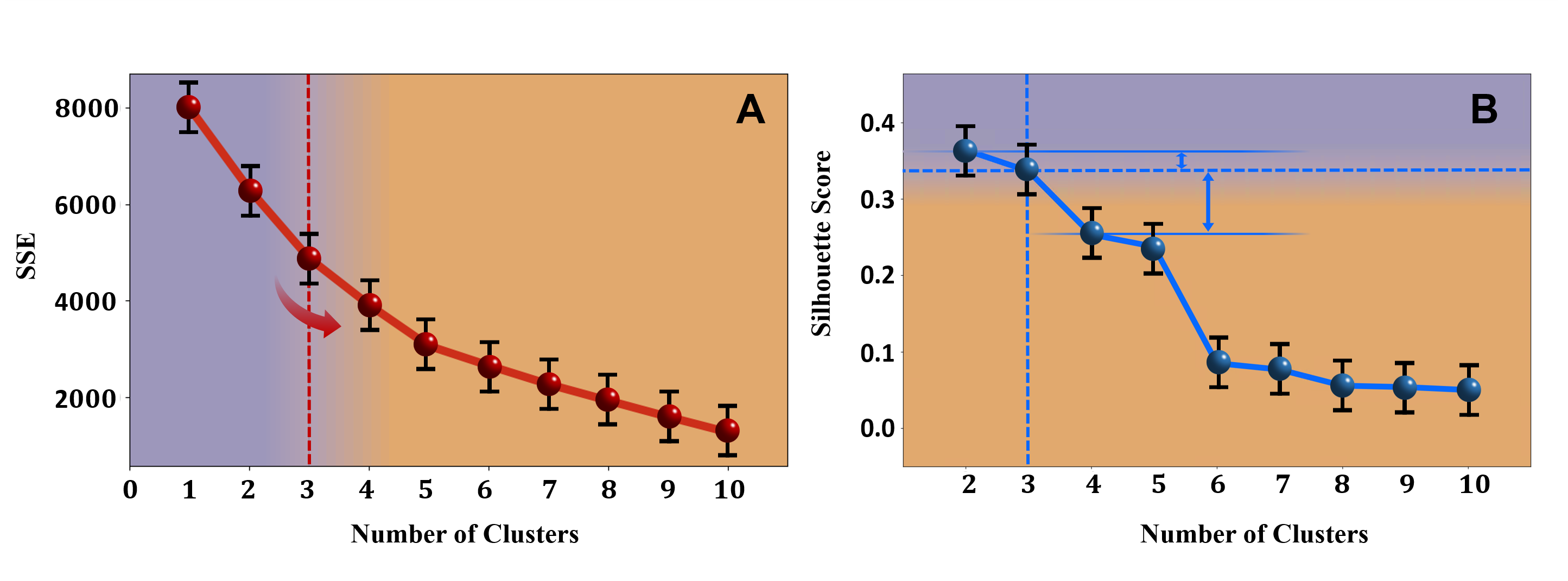}
    \caption{Cluster elbow curve (A) and contour coefficient graph (B)}
    \label{fig:elbow}
\end{figure}

To determine the optimal number of clusters, we create elbow plots and silhouette coefficient graphs, as shown in Figure~\ref{fig:elbow}. The elbow plot offers an intuitive perspective for observing the total variance explained under different numbers of clusters, while the silhouette coefficient assesses the tightness and separation of the clustering. After a comprehensive analysis of both the elbow plot and the silhouette coefficient graph, it becomes evident that a three-cluster solution emerges as the most justifiable categorization approach for the atoms. The elbow plot exhibits a distinct inflection point at three clusters, suggesting a notable reduction in within-cluster variance beyond this point, which diminishes the marginal gain from additional clusters. Concurrently, the silhouette coefficient graph indicates a peak at three clusters, signifying that this number of clusters achieves an optimal balance between cohesion within clusters and separation between them.

\begin{figure}
    \centering
    \includegraphics[width=1.0\linewidth]{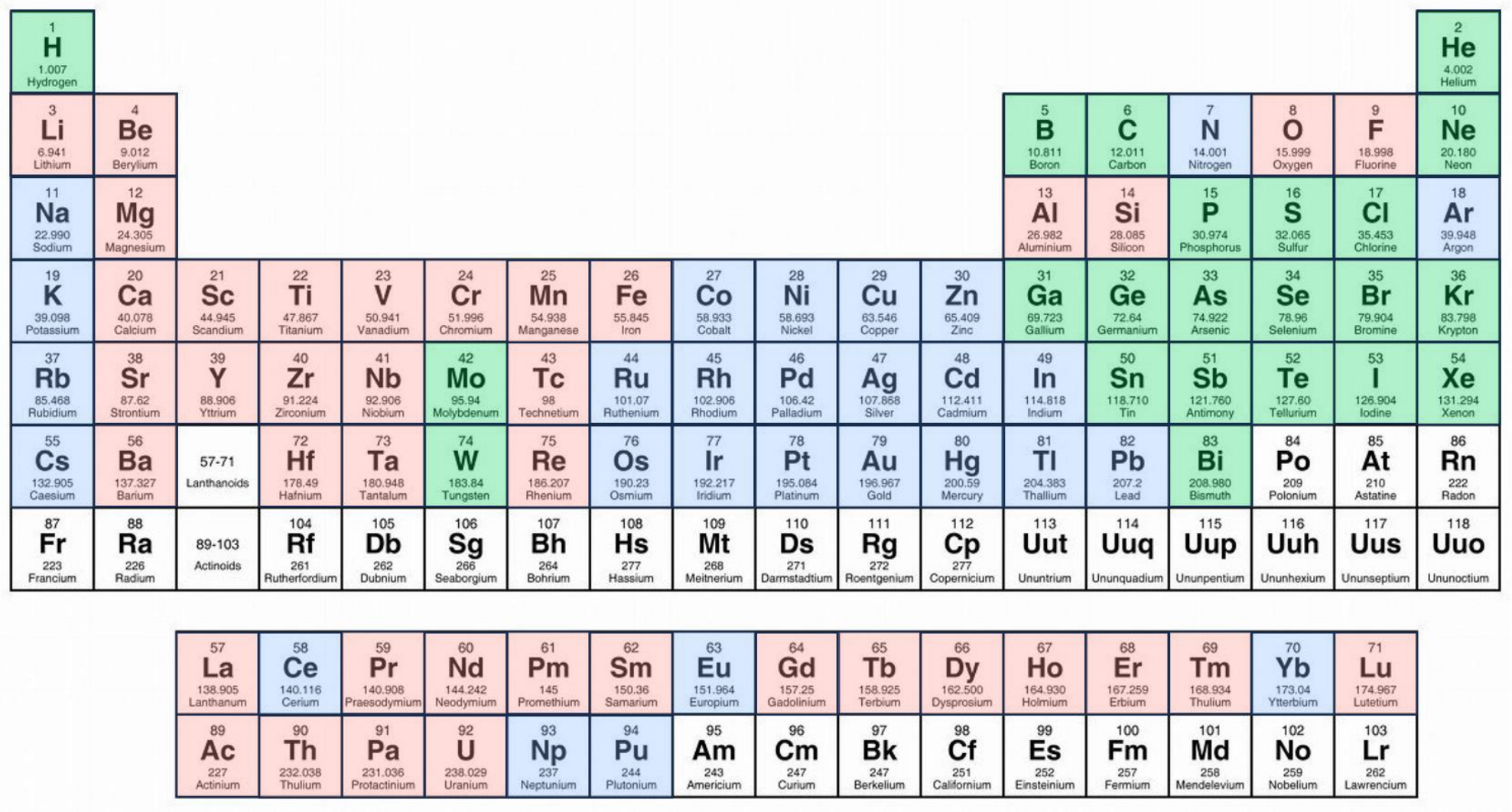}
    \caption{The distribution of three types of elements in the periodic table}
    \label{fig:elements}
\end{figure}

Based on the clustering results, we can categorize the 89 types of atoms into three classes as shown in Figure~\ref{fig:elements}, i.e. 1) \textbf{The first class} encompasses high electronegativity non-metals and some special metal elements, such as boron, carbon, sulfur, gallium, and germanium. These elements play a key role in forming covalent and ionic bonds, significantly influencing the electronic structure and bandgap characteristics of crystals. 2) \textbf{The second class} mainly consists of alkali metals, alkaline earth metals, and some transition metals. These elements typically exist as cations within crystals, crucially affecting the charge balance and structural stability of the crystal. 3) \textbf{The third class} includes inert gases and post-transition metals, such as argon, copper, and zinc. The roles of these elements in crystals vary, ranging from influencing crystal density to participating in complex chemical reactions, thereby directly affecting the properties of the materials.



To further validate the strong physical interpretability of universal atomic embedding and its close connection to the intrinsic properties of atoms, we embark on a further calculation that aims to obtain its accuracy and correlation in the reverse prediction of atomic properties, which involves extracting a series of important elemental properties, including atomic radius, boiling temperature, melting temperature, electrical conductivity, first ionization energy. These properties are selected as dependent variables, while the previously obtained 128-dimensional universal atomic embeddings serve as independent variables in constructing a CatBoost machine learning regression model.

\begin{table}[ht]
\centering
\scriptsize
\caption{Performance comparison of different metrics. The metrics include radius, boiling temperature, melting temperature, electrical conductivity, and first ionization energy.}
\label{tab:metrics_comparison}
\begin{tabular}{cccccc}
\hline
Properties & Radius & Boiling Temperature & Melting Temperature & Electrical Conductivity & First Ionization Energy \\
\hline
$R^2$ & 0.784 & 0.864 & 0.856 & 0.831 & 0.907 \\
\hline
\end{tabular}
\end{table}

The training results were averaged over multiple seeds to ensure the stability and reliability of the outcomes, and the $R^2$ (coefficient of determination) for the model in predicting each property is calculated, with the results listed in Table~\ref{tab:metrics_comparison}.
Analysis of the prediction results reveals that the model exhibits high values of $R^2$, which indicates that, even with small-set data, our universal atomic embeddings are able to establish a robust connection with the physical and chemical properties of atoms.

\begin{figure}[ht!]
    \centering
    \includegraphics[width=0.9\linewidth]{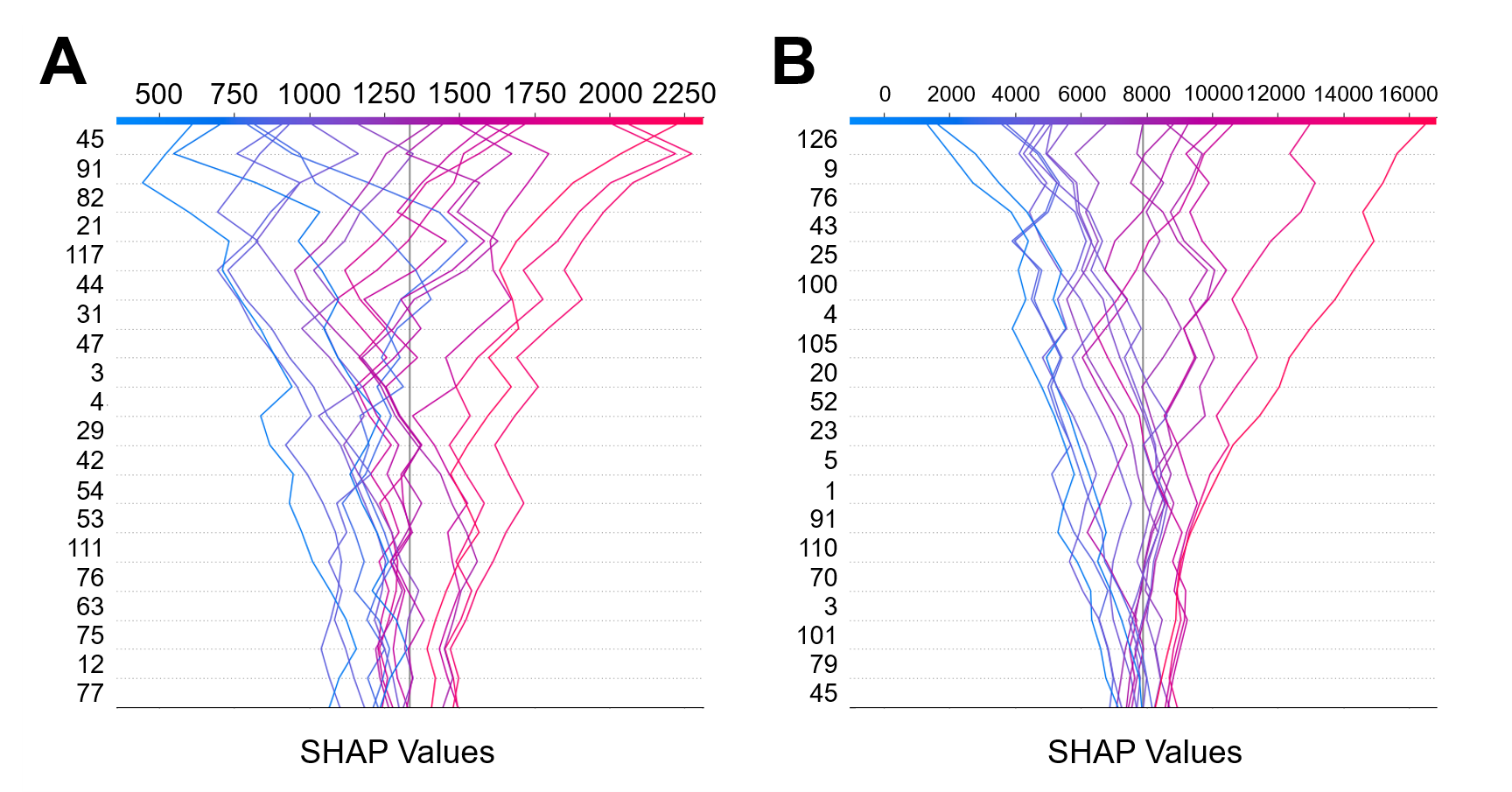}
    \caption{Utilizing the SHAP algorithm to visualize the significance of each dimension within universal atomic embedding for properties such as melting temperature (A) and electrical conductivity (B).}
    \label{fig:shap}
\end{figure}

Furthermore, we employ the SHAP algorithm to visualize the correlation behavior of the 128 feature dimensions of universal atomic embedding in the reverse prediction process of material properties. Melting temperature and electrical conductivity  are selected as examples for visualization in Figure ~\ref{fig:shap},
in which the variables ascending upwards signify increasing importance for the material properties, with the horizontal axis representing the positive and negative correlations. One or more variables of particularly high importance for each material property can be identified and listed in Table~\ref{tab:feature_dimensions}. These variables have a significant impact on the corresponding material properties, exhibiting strong correlations. From this perspective, the feature parameters derived from deep learning networks demonstrate a certain level of interpretability, akin to genes, revealing the fundamental properties and intrinsic connections of materials.

\begin{table}[ht!]
\centering
\caption{Most Important Feature Dimensions for Various Properties}
\label{tab:feature_dimensions}
\begin{tabular}{cc}
\hline
Properties               & Most Important Feature Dimensions \\
\hline
Radius                   & \textbf{98}, 109, 6                        \\
Boiling Temperature      & \textbf{63}, 11                            \\
Melting Temperature      & \textbf{45}, 91, 82, 21                    \\
Electrical Conductivity  & \textbf{126}, 9, 76                        \\
First Ionization Energy  & \textbf{85}, 20, 115                       \\
\hline
\end{tabular}
\end{table}

\subsection{Application in inorganic perovskite dataset}
Inorganic perovskite materials, known for their ABX3 structure, have garnered significant interest in materials science, especially for their potential in solar cell and LED technologies. Their properties, such as high light absorption and tunable bandgaps, position them as strong candidates for next-generation photovoltaic and lighting systems. The integration of perovskite and organic semiconductor polymer subcells into tandem structures is a notable development in this field. These tandem solar cells, combining wide-bandgap perovskite cells and small optical gap organic cells, promise improved device performance and moisture resistance. A notable advancement includes perovskite/organic tandem solar cells achieving a power conversion efficiency (PCE) of over 20\%, with further predictions of exceeding 30\% PCE. This highlights the significant potential of perovskites in renewable energy and sustainable technology\cite{bati2023next}.

However, the research and development of perovskite materials are often impeded by the scarcity of extensive datasets\cite{kim2017hybrid}\cite{nakajima2017discovery}. Unlike more commonly studied materials, perovskites often lack large, comprehensive datasets due to the complexity of their synthesis and characterization.For example,the Hybrid organic-inorganic perovskites (HOIPs) only contains 1346 structures\cite{kim2017hybrid}.This scarcity presents a significant challenge for traditional computational models, which typically rely on extensive data to train and refine their predictions.

Despite these challenges, our application of the CrystalTransformer model's universal atomic embeddings to perovskite datasets marks a substantial stride forward. By merging two distinct datasets of perovskite materials \cite{kim2017hybrid}\cite{nakajima2017discovery}, we created a more diverse and representative dataset which contains 3334 perovskite structures, limited in size compared to datasets for more common materials. Employing a standard Convolutional Graph Neural Network (CGCNN) as our baseline model, we integrated these pre-trained embeddings to enhance the model's predictive accuracy.

The results were remarkable: the mean absolute error (MAE) for predicting formation energy ($E_f$) of perovskite materials was significantly reduced, from 0.054 eV/atom to 0.046 eV/atom, marking a 16\% improvement. This advancement demonstrates not just the efficacy but also the transferability of our pre-trained embeddings in a domain characterized by data scarcity.

Such an improvement is particularly valuable in the field of perovskite research, where the limited availability of data has been a persistent bottleneck. By leveraging the insights gleaned from broader datasets through universal atomic embeddings, we can now make more accurate and reliable predictions even with smaller datasets. This capability is crucial for accelerating the pace of discovery and optimization of new perovskite materials, potentially hastening the advent of more efficient solar cells and LEDs.

The application of universal atomic embeddings in the field of inorganic perovskite materials not only showcases the adaptability and strength of the CrystalTransformer model but also highlights a promising path forward in the pursuit of sustainable and renewable energy solutions. Through such advancements, we are better equipped to address the challenges posed by data scarcity in material science, paving the way for innovative developments in crucial technological domains.

\section{Conclusion}

This study focuses on the application of universal atomic embedding in crystal deep learning, demonstrating its enhancement of overall model capabilities and significant advantages over traditional embedding methods. By mitigating inherent biases in graph neural networks, we introduced the CrystalTransformer model as an universal atomic embedding generator, proving its superior performance in various property tasks. Moreover, rigorous experiments confirmed the exceptional universality and transferability of our universal atomic embedding. We further employed clustering and correlation analyses, along with reverse prediction of atomic properties, to establish our universal atomic embedding as a physically interpretable atomic fingerprint.The application in perovskite datasets highlights the model's capability to make accurate predictions even with limited data, which is pivotal for the rapid discovery and optimization of new perovskite materials. The success demonstrates the effectiveness of our approach in addressing data scarcity.This research also showcases the substantial benefits of universal atomic embedding in molecular simulation, revealing its predictive and analytical prowess in chemical structure analysis. The experimental outcomes offer a new trajectory for deep learning applications in chemistry and material science. Notably, our study identifies areas for improvement, such as the potential role of multitask learning in universal atomic embedding, hindered by the scarcity of crystal property datasets. Future enhancements to the CrystalTransformer model and exploration of fundamental factors influencing universal atomic embedding are envisaged. Despite challenges and limitations, we are confident that universal atomic embedding will play a pivotal role in scientific research and industrial applications.

\section*{Method}

\subsection{Algorithm for the vanilla transformer}\label{sec:transformer}

The transformer architecture\cite{vaswani2017attention}has revolutionized the field of deep learning by providing a mechanism to capture long-range dependencies within sequence data. At its core, the transformer uses stacked self-attention layers and point-wise fully connected layers for both the encoder and decoder. Unlike previous sequence models that required step-wise computation, the transformer processes all elements of the sequence in parallel, leading to substantial gains in efficiency.
Each output element, \( z_i \), in the self-attention mechanism of a transformer is computed as a weighted sum of linearly transformed input elements:

\begin{equation}
z_i = \sum_{j=1}^{n} \alpha_{ij} (x_j W^V),
\end{equation}

where \( \alpha_{ij} \) is the attention weight. These weights are computed using a softmax function to normalize the scores obtained from a scaled dot-product attention mechanism:

\begin{equation}
\alpha_{ij} = \frac{\exp(e_{ij})}{\displaystyle\sum_{k=1}^{n} \exp(e_{ik})},
\end{equation}

The attention scores \( e_{ij} \) themselves are calculated as:

\begin{equation}
e_{ij} = \frac{(x_i W^Q)(x_j W^K)^T}{\sqrt{d_z}},
\end{equation}

where \( d_z \) is the dimensionality of the input embeddings, and \( W^Q \), \( W^K \), and \( W^V \) are parameter matrices that are learned during training. The scaling factor \( \sqrt{d_z} \) is introduced to stabilize the gradient during training, as the dot product can grow large in magnitude, pushing the softmax function into regions where it has extremely small gradients.

One of the key innovations of the transformer is the multi-head attention mechanism, which runs several attention mechanisms in parallel. This design enables the model to capture different types of relationships within the data across different positions and representation subspaces. Furthermore, the incorporation of position encodings imbues the model with sequence order information, which is vital as the self-attention mechanism alone does not consider the order of the input elements.
Additionally, each layer in the transformer includes a residual connection followed by layer normalization, which is critical for training deep architectures. The residual connections allow gradients to flow through the network directly, while layer normalization stabilizes the training process.

\subsection{Multi-task learning method}\label{sec:mt}

MTL\cite{sanyal2018mt} is a learning paradigm in which a model is trained simultaneously on multiple related tasks, sharing representations among them to improve generalization. In the context of the CrystalTransformer model, MTL is instrumental for ensuring the generality of the universal atomic embeddings across various prediction tasks. By leveraging a shared representation, MTL enables the model to generalize better to new tasks or datasets, which is particularly beneficial in the field of material science where data can be scarce.
In defining the loss function for an MTL setup, the total loss is a weighted sum of the losses for each task:

\begin{equation}
\mathcal{L}_{\text{MTL}} = \sum_{i} w_i \cdot \text{Loss}_i(\mathbf{y}_{\text{pred}, i}, \mathbf{y}_{\text{target}, i}),
\end{equation}

where \( \text{Loss}_i \) could be MSE or MAE, \( w_i \) are the task weights, and \( i \) indexes the task. These weights can be set based on the relative importance of each task, the confidence in the task's data, or even learned dynamically during training.
Furthermore, MTL is conducive to ensuring the generality of atomic embeddings, rather than developing an embedding that is overly specialized to a single task. This balance is particularly vital in material science predictive tasks, where embeddings should not only maintain accuracy and efficiency in individual tasks but also generalize well across a range of related tasks. Such a versatile embedding can significantly boost the predictive capabilities and efficiency of the models.

\subsection{Data augmentation for positional invariance}

Data augmentation\cite{fu2020crystallographic} in crystal structure analysis is aimed at training the CrystalTransformer model to recognize invariant properties of crystals. By applying translations, rotations, and reflections — key symmetry operations in crystallography — the training dataset is artificially expanded. This helps the model learn a representation that is consistent with the physical nature of crystals.

\begin{itemize}
\item \textbf{Translation:} Atomic coordinates are translated by a vector \( \mathbf{t} \), a linear combination of the lattice vectors \( \mathbf{a}, \mathbf{b}, \) and \( \mathbf{c} \), simulating the crystal lattice's infinite periodicity:

\begin{equation}
\mathbf{X'} = \mathbf{X} + n_1\mathbf{a} + n_2\mathbf{b} + n_3\mathbf{c},
\end{equation}
where \( n_1, n_2, \) and \( n_3 \) are integers defining the translation in terms of the unit cell parameters.

\item \textbf{Rotation:} Rotation matrices \( \mathbf{R} \) are applied, based on the crystal's point group symmetry, to rotate each structure around a randomly chosen axis:

\begin{equation}
\mathbf{X''} = \mathbf{R}_n \cdot \mathbf{X'},
\end{equation}
where \( \mathbf{R}_n \) is the rotation matrix for the \( n \)-fold rotation axis.

\item \textbf{Reflection:} Reflection matrices corresponding to the crystal's mirror planes are used, reflecting each atomic coordinate across these planes:

\begin{equation}
\mathbf{X'''} = \mathbf{M}_h \cdot \mathbf{X''},
\end{equation}
where \( \mathbf{M}_h \) is the reflection matrix for the mirror plane \( h \).
\end{itemize}

This augmentation strategy ensures that augmented structures remain physically plausible, preserving core chemical and physical properties. It enables the model to adapt to varying data conditions, such as different orientations and positions relative to measurement apparatuses.
The process is calibrated to balance diversity with physical accuracy. Parameters are chosen based on a comprehensive analysis of the crystallographic data distribution and physical characteristics, ensuring that the model learns realistic and relevant patterns without being misled by excessive or irrelevant variations.


\end{document}